\documentstyle[preprint,tighten,prc,aps,epsf]{revtex}

\begin{document}
\draft

\def\bra#1{{\langle#1\vert}}
\def\ket#1{{\vert#1\rangle}}
\def\coeff#1#2{{\scriptstyle{#1\over #2}}}
\def\undertext#1{{$\underline{\hbox{#1}}$}}
\def\hcal#1{{\hbox{\cal #1}}}
\def\sst#1{{\scriptscriptstyle #1}}
\def\eexp#1{{\hbox{e}^{#1}}}
\def\rbra#1{{\langle #1 \vert\!\vert}}
\def\rket#1{{\vert\!\vert #1\rangle}}
\def\lsim{{ <\atop\sim}}
\def\gsim{{ >\atop\sim}}
\def\nubar{{\bar\nu}}
\def\psibar{{\bar\psi}}
\def\Gmu{{G_\mu}}
\def\alr{{A_\sst{LR}}}
\def\wpv{{W^\sst{PV}}}
\def\evec{{\vec e}}
\def\notq{{\not\! q}}
\def\notk{{\not\! k}}
\def\notp{{\not\! p}}
\def\notpp{{\not\! p'}}
\def\notder{{\not\! \partial}}
\def\notcder{{\not\!\! D}}
\def\notA{{\not\!\! A}}
\def\notv{{\not\!\! v}}
\def\Jem{{J_\mu^{em}}}
\def\Jana{{J_{\mu 5}^{anapole}}}
\def\nue{{\nu_e}}
\def\mn{{m_\sst{N}}}
\def\mns{{m^2_\sst{N}}}
\def\me{{m_e}}
\def\mes{{m^2_e}}
\def\mq{{m_q}}
\def\mqs{{m_q^2}}
\def\mz{{M_\sst{Z}}}
\def\mzs{{M^2_\sst{Z}}}
\def\ubar{{\bar u}}
\def\dbar{{\bar d}}
\def\sbar{{\bar s}}
\def\qbar{{\bar q}}
\def\sstw{{\sin^2\theta_\sst{W}}}
\def\gv{{g_\sst{V}}}
\def\ga{{g_\sst{A}}}
\def\pv{{\vec p}}
\def\pvs{{{\vec p}^{\>2}}}
\def\ppv{{{\vec p}^{\>\prime}}}
\def\ppvs{{{\vec p}^{\>\prime\>2}}}
\def\qv{{\vec q}}
\def\qvs{{{\vec q}^{\>2}}}
\def\xv{{\vec x}}
\def\xpv{{{\vec x}^{\>\prime}}}
\def\yv{{\vec y}}
\def\tauv{{\vec\tau}}
\def\sigv{{\vec\sigma}}
\def\sst#1{{\scriptscriptstyle #1}}
\def\gpnn{{g_{\sst{NN}\pi}}}
\def\grnn{{g_{\sst{NN}\rho}}}
\def\gnnm{{g_\sst{NNM}}}
\def\hnnm{{h_\sst{NNM}}}

\def\xivz{{\xi_\sst{V}^{(0)}}}
\def\xivt{{\xi_\sst{V}^{(3)}}}
\def\xive{{\xi_\sst{V}^{(8)}}}
\def\xiaz{{\xi_\sst{A}^{(0)}}}
\def\xiat{{\xi_\sst{A}^{(3)}}}
\def\xiae{{\xi_\sst{A}^{(8)}}}
\def\xivtez{{\xi_\sst{V}^{T=0}}}
\def\xivteo{{\xi_\sst{V}^{T=1}}}
\def\xiatez{{\xi_\sst{A}^{T=0}}}
\def\xiateo{{\xi_\sst{A}^{T=1}}}
\def\xiva{{\xi_\sst{V,A}}}

\def\rvz{{R_\sst{V}^{(0)}}}
\def\rvt{{R_\sst{V}^{(3)}}}
\def\rve{{R_\sst{V}^{(8)}}}
\def\raz{{R_\sst{A}^{(0)}}}
\def\rat{{R_\sst{A}^{(3)}}}
\def\rae{{R_\sst{A}^{(8)}}}
\def\rvtez{{R_\sst{V}^{T=0}}}
\def\rvteo{{R_\sst{V}^{T=1}}}
\def\ratez{{R_\sst{A}^{T=0}}}
\def\rateo{{R_\sst{A}^{T=1}}}

\def\mro{{m_\rho}}
\def\mks{{m_\sst{K}^2}}
\def\mpi{{m_\pi}}
\def\mpis{{m_\pi^2}}
\def\mom{{m_\omega}}
\def\mphi{{m_\phi}}
\def\Qhat{{\hat Q}}

\def\FOS{{F_1^{(s)}}}
\def\FTS{{F_2^{(s)}}}
\def\GAS{{G_\sst{A}^{(s)}}}
\def\GES{{G_\sst{E}^{(s)}}}
\def\GMS{{G_\sst{M}^{(s)}}}
\def\GATEZ{{G_\sst{A}^{\sst{T}=0}}}
\def\GATEO{{G_\sst{A}^{\sst{T}=1}}}
\def\mdax{{M_\sst{A}}}
\def\mustr{{\mu_s}}
\def\rsstr{{r^2_s}}
\def\rhostr{{\rho_s}}
\def\GEG{{G_\sst{E}^\gamma}}
\def\GEZ{{G_\sst{E}^\sst{Z}}}
\def\GMG{{G_\sst{M}^\gamma}}
\def\GMZ{{G_\sst{M}^\sst{Z}}}
\def\GEn{{G_\sst{E}^n}}
\def\GEp{{G_\sst{E}^p}}
\def\GMn{{G_\sst{M}^n}}
\def\GMp{{G_\sst{M}^p}}
\def\GAp{{G_\sst{A}^p}}
\def\GAn{{G_\sst{A}^n}}
\def\GA{{G_\sst{A}}}
\def\GETEZ{{G_\sst{E}^{\sst{T}=0}}}
\def\GETEO{{G_\sst{E}^{\sst{T}=1}}}
\def\GMTEZ{{G_\sst{M}^{\sst{T}=0}}}
\def\GMTEO{{G_\sst{M}^{\sst{T}=1}}}
\def\lamd{{\lambda_\sst{D}^\sst{V}}}
\def\lamn{{\lambda_n}}
\def\lams{{\lambda_\sst{E}^{(s)}}}
\def\bvz{{\beta_\sst{V}^0}}
\def\bvo{{\beta_\sst{V}^1}}
\def\Gdip{{G_\sst{D}^\sst{V}}}
\def\GdipA{{G_\sst{D}^\sst{A}}}
\def\fks{{F_\sst{K}^{(s)}}}
\def\FIS{{F_i^{(s)}}}
\def\fpi{{F_\pi}}
\def\fk{{F_\sst{K}}}

\def\RAp{{R_\sst{A}^p}}
\def\RAn{{R_\sst{A}^n}}
\def\RVp{{R_\sst{V}^p}}
\def\RVn{{R_\sst{V}^n}}
\def\rva{{R_\sst{V,A}}}
\def\xbb{{x_B}}

\def\lamtv{{\Lambda_\sst{TVPC}}}
\def\lamtvs{{\Lambda_\sst{TVPC}^2}}
\def\lamtvc{{\Lambda_\sst{TVPC}^3}}
\def\lamtvi{{\Lambda_\sst{TVPC}^{-1}}}
\def\lampv{{\Lambda_\sst{PV}}}

\def\osffp{{{\cal O}_7^{ff'}}}
\def\ospg{{{\cal O}_7^{\gamma g}}}
\def\ospz{{{\cal O}_7^{\gamma Z}}}

\def\PR#1{{{\em   Phys. Rev.} {\bf #1} }}
\def\PRC#1{{{\em   Phys. Rev.} {\bf C#1} }}
\def\PRD#1{{{\em   Phys. Rev.} {\bf D#1} }}
\def\PRL#1{{{\em   Phys. Rev. Lett.} {\bf #1} }}
\def\NPA#1{{{\em   Nucl. Phys.} {\bf A#1} }}
\def\NPB#1{{{\em   Nucl. Phys.} {\bf B#1} }}
\def\AoP#1{{{\em   Ann. of Phys.} {\bf #1} }}
\def\PRp#1{{{\em   Phys. Reports} {\bf #1} }}
\def\PLB#1{{{\em   Phys. Lett.} {\bf B#1} }}
\def\ZPA#1{{{\em   Z. f\"ur Phys.} {\bf A#1} }}
\def\ZPC#1{{{\em   Z. f\"ur Phys.} {\bf C#1} }}
\def\etal{{{\em   et al.}}}

\def\delalr{{{delta\alr\over\alr}}}
\def\pbar{{\bar{p}}}
\def\lamchi{{\Lambda_\chi}}
\def\gaf{{g_\sst{A}^f}}
\def\gvu{{g_\sst{V}^u}}
\def\gvd{{g_\sst{V}^d}}
\def\gve{{g_\sst{V}^e}}

\title{
Electric Dipole Moments and the Mass Scale of\\
New T-Violating, P-Conserving Interactions}

\author{
M.J. Ramsey-Musolf
\thanks{National Science Foundation
Young Investigator}\\[0.3cm]
}
\address{
Department of Physics, University of Connecticut\\
Storrs, CT 06269 USA\\
\ \\
and
 \ \\
Theory Group, Thomas Jefferson National Accelerator Facility\\
Newport News, VA 23606 USA}


\maketitle

\begin{abstract}
We consider the implications of experimental limits on the permanent
electric dipole moment (EDM) of
the electron and neutron for possible new  time-reversal violating
(TV)  parity-conserving (PC) interactions. We show that the constraints
derived from one-loop
contributions to the EDM exceed previously reported two-loop limits by more
than an order of
magnitude and imply a lower bound on the new TVPC mass scale $\lamtv$ of
150 TeV for new TVPC strong
interactions. These results imply a value of $10^{-15}$ or smaller for the
ratio of low-energy TVPC
matrix elements to those of the residual strong interaction.
\end{abstract}


\narrowtext

The search for physics beyond the Standard Model is a topic of on-going
interest for both high-energy colliders as well as low-energy experiments
involving atoms and nuclei. Although the Standard Model is enormously
successful
in accounting for a plethora of electroweak data of a broad range in energies,
there exist strong theoretical reasons for considering the Standard Model as an
effective theory -- derived from some broader framework -- applicable to
physics
below the weak scale. Among the questions to be addressed in considering
possible
frameworks is the mass scale $\Lambda$ associated with \lq\lq new physics". In
this respect, experiments in atomic parity violation (APV) provide a powerful
probe of new physics scenarios which violate parity. Recently, the Boulder
Group has used
APV to determine the weak charge $Q_W$ of the cesium atom\cite{Ben99}. The
reported value for
$Q_W$, which differs from the Standard Model prediction by 1.5\%
(2.5$\sigma$), implies
the existence of new PV interactions with mass scales $\lampv$ on the order
of 1 TeV or
greater\cite{MRM99}.

In this Letter, we consider $\lamtv$, the mass scale associated with
possible new
time-reversal violating (TV), parity-conserving (PC) interactions. Explicit
searches
for new TVPC effects at low-energies have been carried out using studies of
detailed
balance in nuclear reactions\cite{Hax94} and neutron transmission
experiments\cite{Kos91}. These
studies imply that $\alpha_T\lsim $ few $\times 10^{-3}$, where $\alpha_T$
gives the ratio
of typical TVPC nuclear matrix elements to those of the residual strong
interaction. The
corresponding limits on the TVPC mass scale are weak: $\lamtv\gsim 10$ GeV.
As we argue
below, significantly more stringent limits can be inferred indirectly from
searches for a
permanent electric dipole moment of the electron and neutron.

It has been pointed out in a series of recent papers that the
lowest-dimension flavor
conserving TVPC interactions have dimension seven\cite{Con92,Eng96}. Such
interactions can generate
a permanent electric dipole moment (EDM) of an elementary fermion or its
many-body
bound states in the presence of a PV Standard Model radiative correction.
It was
argued in these studies that the most restrictive limit on new dimension seven
TVPC interactions is obtained from a two-loop contribution to the EDM, and the
expected magnitude of low-energy TVPC observables was inferred. No attempt was
made to derive a lower bound on $\lamtv$. In what follows, we show that there
exist additional $d=7$ operators not considered previously which contribute
to the
EDM at one-loop order and which generate more stringent lower bounds on
$\lamtv$ than
those derived at two-loop order. We also revisit the analysis of Ref.
\cite{Con92} and
argue that it is inconsistent with the separation of scales and systematic
power counting which
underlies low-energy effective field theory (LEEFT). LEEFT is the
appropriate framework for
analyzing the non-renormalizable interactions of interest here. The
corresponding scale separation,
which is preserved when loop integrals are regulated using dimensional
regularization (DR), implies a
different $\lamtv$-dependence for the EDM than obtained in Ref.
\cite{Con92}. Using an explicit
calculation, we obtain the correct scaling of the EDM with $\lamtv$ and
derive lower bounds on
$\lamtv$ under naturalness assumptions for the coefficients of the $d=7$
TVPC operators.
We find these bounds are significantly stronger than the scale $\lampv$
obtained from APV.
Our results also imply $\alpha_T\sim 10^{-15}$ or smaller, independent of
any naturalness
assumptions.

Although the origins of possible new TVPC interactions are not known, it
has been shown
that they cannot arise via tree-level boson exchange in a renormalizable
gauge theory\cite{Her92}.
Hence, one  expects them to be generated either by loop effects or
non-perturbative short distance
dynamics. Consequently, it is convenient to describe
its low-energy consequences using effective Lagrangians. Following Ref.
\cite{Eng96}, we write
\begin{equation}
\label{lnew}
{\cal L}_\sst{NEW} = {\cal L}_4 +{1\over\lamtv}{\cal
L}_5+{1\over\lamtvs}{\cal L}_6
+{1\over\lamtvc}{\cal L}_7+\cdots \ \ \ ,
\end{equation}
 where the subscripts denote operator dimension. The TVPV EDM operator
appears in ${\cal L}_5$:
\begin{equation}
{\cal O}_5 = -\frac{i}{2} C_5^f {\bar\psi}\sigma_{\mu\nu}\gamma_5\psi\
F^{\mu\nu}\ \ \ .
\end{equation}
The TVPC operators considered in Refs. \cite{Con92,Eng96} appear in ${\cal
L}_7$:
\begin{eqnarray}
{\cal O}_7^{ff'}&=&
C_7^{ff'} {\bar\psi}_f {\buildrel \leftrightarrow \over D_\mu} \gamma_5 \psi_f
	   {\bar\psi}_{f'}\gamma^\mu\gamma_5 \psi_{f'}\\
{\cal O}_7^{\gamma g} &=& C_7^{\gamma g} {\bar\psi}\sigma_{\mu\nu}\lambda^a\psi
    F^{\mu\lambda} G_\lambda^{a\ \nu}\ \ \ ,
\end{eqnarray}
where $f$ and $f'$ are distinct fermions and $F_{\mu\nu}$ and
$G^a_{\mu\nu}$ are the
photon and gluon field strength tensors, respectively. Both ${\cal
O}_7^{ff'}$ and
${\cal O}_7^{\gamma g}$ contribute to $C_5^f$ at two-loop order, although
only the
contribution of the four-fermion interaction has been computed explicitly
previously.
In addition, we consider the following TVPC operator appearing in ${\cal L}_7$:
\begin{equation}
{\cal O}_7^{\gamma Z} = C_7^{\gamma Z} {\bar \psi} \sigma_{\mu\nu}\psi\
F^{\mu\lambda}
     Z_\lambda^\nu \ \ \ ,
\end{equation}
where $Z_{\mu\nu}$ is the $Z$-boson field strength tensor. As we show
below, ${\cal
O}_7^{\gamma Z}$ contributes to $C_5^f$ at one-loop order and yields the
strongest bound on
$\lamtv$.

Implicit in the LEEFT expansion of ${\cal L}_\sst{NEW}$ in Eq. (\ref{lnew})
is a separation of
scales.  Short distance effects ($\mu\gsim\lamtv$) are subsumed into the
renormalized operator
coefficients $C_{n}$. In the present case, these short distance effects are
not calculable, since
the full theory for $\mu\gsim\lamtv$ is unknown. Long distance
contributions ($\mu\lsim\lamtv$)
arise from matrix elements of the effective operators ${\cal O}_n$ taken
between states containing
only particles having masses and momenta below $\lamtv$. When these matrix
elements involve
divergent loops containing the ${\cal O}_n$, the use of  DR and $\overline{MS}$ renormalization
preserves the LEEFT separation of scales by protectin
g loop integrals from high momentum
($p\gsim\lamtv$) contributions. Preservation of the scale separation is
critical to maintaining
the power counting (in $1/\lamtv$) associated with Eq. (\ref{lnew}).
Moreover, it implies that
renormalization of ${\cal O}_5$ from loops containing the ${\cal O}_7$ must
scale as $(M/\lamtv)^2$,
where $M$ is the mass of one of the particles dynamically relevant for
$\mu\lsim\lamtv$.

We observe that if the parity symmetry broken by the SM is not restored for
scales $\mu\gsim\lamtv$,
then the coefficient $C_5$ must exist at tree-level in the LEEFT. Since
both the SM PV interaction
and the fundamental, but not calculable, interactions which generate the
$d=7$ TVPC interactions
exist at such scales, there exist no reason for them not to conspire in
generating a non-vanishing
$C_5$. In this case, power counting implies that loops containing the
${\cal O}_7$ will generate
subdominant contributions to $C_5$ (see below), so that the EDM limits
cannot be used to constrain
$d=7$ TVPC operators. At best, one may employ dimensional arguments
involving $C_5$ to derived
lower bounds on $\lamtv$. For example, taking $C_5=4\pi\kappa^2 e$ and
using the present limits on
the electron EDM, one obtains $\lamtv\gsim 10^{14} \kappa^2$ GeV. This
bound is considerably stronger
than obtained by the authors of Ref. \cite{Con92}, who presume,
incorrectly, to be able to calculate
short-distance effects via loops.

A more interesting scenario occurs when parity symmetry is restored above
the weak scale but below
$\lamtv$ ({\em e.g.}, in a left-right symmetric scenario). In this case,
$C_5=0$ at tree-level in the
LEEFT and becomes non-vanishing only through PV radiative corrections to
the $d=7$ (and higher) TVPC
interactions. A conservative lower bound on $\lamtv$ can be obtained by
considering the SM PV
radiative corrections.  The leading order contributions to the fermion EDM
arising from the TVPC
operators in
${\cal L}_7$ arise from the diagrams of Figs. 1 and 2. For simplicity, we
consider only the
effects of $\ospz$ (Fig. 1) and $\osffp$ (Fig. 2). The conclusions obtained
from the
two-loop gluon-$Z$ graphs will be similar. Following Refs.
\cite{Con92,Eng96}, we also restrict our
attention to neutral current PV corrections. The diagrams diverge
quadratically. Following the
standard practice of LEEFT, we regulate the integrals using DR
and subtract the pole terms in the $\overline{MS}$ scheme with the
appropriate counterterm in
$C_5^f$. In the case of Fig. 1, the PV effect arises from the axial vector
$Z$-fermion coupling. In
the leading log-approximation, the resulting finite contribution to the EDM
from $\ospz$ is
\begin{equation}
\label{oneloop}
C_5^f\sim e C_7^{\gamma Z} \left({\mz\over\lamtv}\right)^2\left({1\over s_W
c_W}\right)
   \gaf\left({1\over 96\pi^2}\right) \ln{\mzs\over\mu^2} \ \ \ ,
\end{equation}
where we have dropped terms quadratic in the fermion mass, where $\gaf$ is
the axial
vector $Zff$ coupling, and where $s_W=\sin\theta_W$ is the sine of the
Weinberg angle.

In the case of the two-loop contribution generated by $\osffp$, the
dominant terms arise
arise from the graphs appearing in Fig. 2. All other two-loop contributions
containing this
operator are suppressed by powers of $m_f/\mz$ where $m_f$ is a (light)
fermion mass.
Turning first to the graphs of Fig. 2a, we note that  the closed fermion
loop containing three
insertions is  identical to the triangle graph appearing in the ABJ
anomaly. Here,
the vector current insertions are associated with the neutral gauge bosons
and the axial
vector insertion arises from $\osffp$.
Denoting its nominally linearly-divergent amplitude $T^{\mu\lambda\alpha}$,
we choose the loop
momentum routing to satisfy $q^\mu T_{\mu\lambda\alpha}=0=k^\lambda
T_{\mu\lambda\alpha}$, where
$q_\mu$ and $k_\lambda$ are the photon and $Z$-boson momenta, respectively.
The result is finite. We have verified that our result produces the
textbook result for
$(q+k)^\alpha T_{\mu\lambda\alpha}$ for $k^2=q^2=0$ \cite{IandZ}.

The remaining integration for the two-loop amplitude of Fig. 2a is
straightforward. Since the
amplitude contains no infrared singularities, we follow Ref. \cite{Con92}
and neglect the
$m_{f'}$-dependence of $T^{\mu\lambda\alpha}$. As with the amplitude for
Fig 1, the two-loop
amplitude diverges quadratically, and we follow the same subtraction
procedure as in the one-loop
case. The corresponding, leading-log finite contribution to the EDM is
\begin{equation}
\label{twoloop}
C_5^f\sim - e C_7^{ff'} \left({\mz\over\lamtv}\right)^2
Q_{f'}g_\sst{V}^{f'}\gaf \left({G_\sst{F}
   \mzs\over\sqrt{2}}\right)\left({1\over 8\pi^2}\right)^2
\ln{\mzs\over\mu^2}\ \ \ ,
\end{equation}
where $g_\sst{V}^{f'}$ ($\gaf$) is the vector (axial vector) coupling of
the $Z$ to fermion
$f'$ ($f$) and $Q_{f'}$ is the EM charge of fermion $f'$, with $f'$
denoting the species of fermion
in the closed loop.

In the case of Fig. 2b, the closed fermion loop contains the axial vector
$Z$-fermion insertion,
while the external fermion couples to the $Z$ through the vector current.
The closed fermion loop
sub-graph diverges quadratically and must be renormalized by the
appropriate $\overline{MS}$
counterterm before the second loop integration is carried out. These graphs
receive contributions
from both a photon insertion on the line for fermion $f$ (the external
fermion) as well as from
the EM seagull vertex generated by the covariant derivative in $\osffp$. To
leading-log order, the
sum of the amplitudes for Fig. 2b gives
\begin{equation}
\label{compton}
C_5^f\sim  - e C_7^{ff'} \left({5\over
12}\right)\left({\mz\over\lamtv}\right)^2 Q_f g_\sst{V}^f
g_\sst{A}^{f'} \left({G_\sst{F}\mzs\over\sqrt{2}}\right)\left({1\over
8\pi^2}\right)^2
\left(\ln{\mzs\over\mu^2}\right)^2\ \ \ ,
\end{equation}
where as before $f'$ is the fermion in the closed loop. The appearance of
the $\ln^2$ arises from
the presence of two sub-divergences in the graphs of Fig. 2b whereas the
closed fermion loop
in Fig. 2a is finite.  For $\ln(\mzs/\mu^2)\sim 1$, the contribution
in Eq. (\ref{compton}) will be of the same order as that in Eq.
(\ref{twoloop}). As we argue below,
however, we expect $\ln(\mzs/\mu^2)\sim 10$, so that the graphs in Fig. 2b
generally give a somewhat
larger contribution that those of Fig. 2a.

We emphasize that the $\lamtv$-dependence appearing in Eqs.
(\ref{oneloop}-\ref{compton}) differs
substantially from that obtained in Refs. \cite{Con92,Eng96}. The reason is
that the regulator used
in the calculation of Ref. \cite{Con92} does not preserve the LEEFT
separation of scales and
effectively mixes all orders in the $1/\lamtv$ expansion. The two-loop
integral containing $\osffp$
was regulated by assuming this interaction arises from the exchange of a
hypothetical axial vector
boson of mass $\lamtv$ having a non-renormalizable coupling to fermion $f$.
The presence of the axial
vector propagator renders the loop integral finite. In effect, this
propagator functions as a form
factor $(p^2/\lamtvs-1)^{-1}$ which contains an infinite power series in
$(p/\lamtv)^2$ with
pre-determined (model-dependent) coefficients. To be consistent, the
effects of an infinite tower of
higher-dimension operators in ${\cal L}_\sst{NEW}$ must also be included in
tandem with this form
factor, though as a practical matter this was not done in the calculation
of Ref. \cite{Con92}.
Furthermore, each operator in the tower will generate an equally important
contribution to the EDM,
and truncation at $d=7$ will be unjustified.

This loss of power counting incurred by form factors can be seen in the
following example.
Consider the tower of operators
\begin{equation}
\label{tower}
{\cal O}_{7+2n} = C^{ff'}_{7+2n} {\bar\psi}_f {\buildrel \leftrightarrow
\over D_\mu}
\gamma_5 \psi_f (\partial^2)^n
	   {\bar\psi}_{f'}\gamma^\mu\gamma_5 \psi_{f'}\ \ \ ,
\end{equation}
where $n=0,1,\ldots$. Inserting these operators into the loops of Fig. 2
generates divergent
contributions to $C_5$. Following the spirit of Ref. \cite{Con92}, we may
regulate the integrals
by including the form factors $(p^2/\lamtvs-1)^{-(n+1)}$. Doing so is
equivalent to repeating the
calculation of Ref. \cite{Con92} with additional factors of
$(p^2/\lamtvs)^n(p^2/\lamtvs-1)^{-n}=1+\cdots$ in the loop integrals. The
first term of order
unity  generates the same leading-log contribution as given in Ref.
\cite{Con92}, while the
remaining terms ($+\cdots$) generate finite contributions for
$\lamtv\to\infty$. At leading-log
order, then, the contribution from the entire tower of operators in Eq.
(\ref{tower}) is
proportional to $\sum_{n=0}^\infty C^{ff'}_{7+2n}$. A similar conclusion
follows if a different form
factor is used to cut the integrals off at $p\sim\lamtv$; each of the
$C^{ff'}_{7+2n}$ will
contribute with a similar weight. Consequently, no information about the
$d=7$ interactions alone can
be extracted from EDM limits.

In effect, the use of a form factor as in Ref. \cite{Con92} allows
contributions from intermediate
states having momenta $p\sim\lamtv$, thereby blurring the separation of
scales implicit in the
low-energy expansion of Eq. (\ref{lnew}). Consequently, the renormalization
of ${\cal O}_5$ due to
any $d\geq 7$ operator is dominated by these high-momentum intermediate
states -- a feature reflected
by the absence of the factors $(\mz/\lamtv)^2$ in the expressions of Ref.
\cite{Con92}. One
therefore has no systematic power counting to justify truncation at $d=7$
in the expansion of Eq.
(\ref{lnew}). In contrast, the use of DR and $\overline{MS}$ subtraction as
above avoids these
high-mass contributions and maintains the power counting in $\lamtvi$
appropriate to the LEEFT
separation of scales.

>From the one- and two-loop results of Eqs. (\ref{oneloop}-\ref{compton})
and the
experimental limits on EDM's, one may derive conservative lower-bounds on
$\lamtv$. In the case of
$\osffp$ contributions, one must specify the fermion species $f'$ involved
in the closed fermion
loop. Since the result in Eq. (\ref{twoloop}) is proportional to
$g_\sst{V}^{f'}$, contributions
involving closed, charged lepton loops are suppressed by
$g_\sst{V}^{\ell^-}=-1+4\sstw\approx
0.1$ with respect to quark loop contributions. Consequently, we consider
only the latter. In this
case,  the constants relevant to the electron and neutron EDM's are
$C_7^{eu}$, $C_7^{ed}$, and
$C_7^{ud}$. Moreover, since $|\gvu|$ and $|\gvd|$ differ by less than a
factor of two, and since
the contributions from $\osffp$ to the EDM go as $1/\lamtvc$, the lower
bounds on $\lamtv$ from
$u$-quark and $d$-quark loops differ negligibly. For the results in Eq.
({\ref{compton}), the
additional factor of $\ln(\mzs/\mu^2)$ renders the contribution from Fig.
2b comparable in magnitude
to that of Fig. 2a when $\mu$ is chose as discussed below.

In the case of the neutron EDM, we use the quark model to relate $d_n$ to
the light quark EDM's.
Following the procedures of Ref. \cite{DGH86}, we obtain
\begin{equation}
\label{quarkmodel}
d_n={1\over\lamtv}\int\ d^3x (u^2+\frac{1}{3}\ell^2)\ \left[\frac{4}{3} C_5^d -
\frac{1}{3} C_5^u\right]\ \ \ ,
\end{equation}
where $u$ and $\ell$ and the upper and lower component quark model radial
wavefunctions,
respectively. Using the wavefunction normalization condition $\int\ d^3x
(u^2+\ell^2) = 1$
and expression for the axial vector charge $\int\ d^3x
(u^2-\frac{1}{3}\ell^2)=\frac{3}{5}g_\sst{A}$
we obtain a value of $(1/4)[1+6g_\sst{A}/5]\approx 0.63$ for the integral
in Eq.
(\ref{quarkmodel}).

The use of Eqs. (\ref{oneloop}-\ref{compton}) to derive limits on $\lamtv$
requires a choice of
renormalization scale $\mu$ and assumptions regarding the constants $C_7$.
Since the typical momentum of a quark inside a nucleon is
$\sim\Lambda_\sst{QCD}$, we take
$\mu=\Lambda_\sst{QCD}$ for $d_n$. The precise choice for this scale does
not affect the lower bounds
on $\lamtv$ appreciably, since it enters only logarithmically.
Consequently, we use the same choice
for the electron EDM, though a smaller scale is likely more appropriate.
The lower bounds on
$\lamtv$ are similarly rather insensitive to the value of the constants
$C_7$ assuming they fall
within a natural range.
Following common conventions \cite{MRM99}, we write $C_7^{ff'} =
4\pi\kappa^2$,
where $\kappa$ specifies the coupling strength of the new TVPC interaction
($\kappa^2\sim 1$ for
new strong interactions).
We also take $ C_7^{\gamma Z} \sim e g C_7^{ff'} =
{4\pi\alpha/\sin\theta_\sst{W}}C_7^{ff'}$ ,
since one would expect $C_7^{\gamma Z}$ to be suppressed by with respect to
$C_7^{ff'}$ by the gauge
couplings associated with the $\gamma$ and $Z$.
With these conventions, we obtain the following lower limits on $\lamtv$
from the experimental
result $|d_e|< 4\times 10^{-27}\ e-{\hbox{cm}}$ \cite{Com94}:
$\lamtv\gsim 150 \kappa^{2/3}$ TeV from the one loop graph of Fig. 1 and
$\lamtv\gsim 30 \kappa^{2/3}$ TeV
from the two-loop graphs of Fig. 2. The corresponding bounds from the
neutron EDM are somewhat weaker
 -- given that the experimental limit on $|d_n|$ is an order of magnitude
larger than the limit on
$|d_e|$ \cite{Alt92}.



Finally, we note the implications of the EDM results for low-energy
measurements of TVPC
observables. As argued on dimensional grounds in Ref. \cite{Eng96}, the
ratio $\alpha_T$
should scale as $C_7\ (p/\lamtv)^3$,
where $p$ is a typical momentum involved in low-energy hadronic
interactions. The experimental EDM
limits constrain the ratio $C_7/\lamtvc\propto \kappa^2/\lamtvc$ as
discussed above. Conservatively
taking $p=1$ GeV$/c$\footnote{Low-energy hadronic interactions are
typically characterized by
momentum transfers of one GeV$/c$ or less}, the our results imply that
$\alpha_T$ should be of the
order of $10^{-15}$ or smaller, independent of the choice of $\kappa$.
Presently, direct TVPC
measurements -- such as compound nucleus studies of detailed balance and
neutron transmission
experiments --  yield limits of about $10^{-3}$ for $\alpha_T$ \cite{Hax94}.

\acknowledgements

It is a pleasure to thank G. Dunne, W. Haxton, and M. Savage for useful
discussions. This work was
supported in part by a National Science Foundation Young Investigator award.

\begin{figure}
\caption{\label{Fig1}One-loop contributions to  EDM of elementary fermion
$f$. The $\otimes$ denotes
the operator $\ospz$.}
\end{figure}

\begin{figure}
\caption{\label{Fig2}Two-loop contributions containing $\osffp$
(denoted by $\otimes$) to the EDM of elementary fermion $f$.}
\end{figure}

\end{document}